\documentclass{PoS}

\usepackage{booktabs,xspace,url,verbatim,amsmath}

\title{Using Muon Rings for the Optical Throughput Calibration of the Cherenkov Telescope Array}

\ShortTitle{Muon Calibration for CTA}

\author{Markus Gaug,$^a$ Stephen Fegan,$^b$ \speaker{Alison Mitchell},$^c$ Maria-Concetta Maccarone,$^d$ Teresa Mineo,$^d$ and Akira Okumura$^e$ for the CTA Consortium\footnote{for consortium list see PoS(ICRC2019)1177}\\
\llap{$^a$}Unitat de F\'isica de les Radiacions, Departament de F\'isica, and CERES-IEEC, Universitat Aut\`onoma de Barcelona, E-08193 Bellaterra, Spain\\
\llap{$^b$}Laboratoire Leprince-Ringuet, Ecole Polyt\'echnique, CNRS/IN2P3, F-91128 Palaiseau, France\\
\llap{$^c$}Physik Institut, Universit\"at Z\"urich, Winterthurerstrasse 190, CH-8057 Z\"urich, Switzerland  \\
\llap{$^d$}Istituto di Astrofisica Spaziale e Fisica Cosmica di Palermo, INAF, Via Ugo La Malfa 153, I-90146 Palermo, Italy\\
\llap{$^e$}Institute for Space–Earth Environmental Research, and 
Kobayashi–Maskawa Institute for the Origin of Particles and the Universe, Nagoya University, Furo-cho, Chikusa-ku, Nagoya 464-8602, Japan\\
E-mail: \email{markus.gaug@uab.cat}, \email{sfegan@llr.in2p3.fr}, \email{alison.mitchell@physik.uzh.ch}, \email{cettina.maccarone@inaf.it}, \email{teresa.mineo@inaf.it}, \email{oxon@mac.com}}

\abstract{Muon ring images observed with Imaging Atmospheric Cherenkov Telescopes (IACTs) provide a powerful means to calibrate the optical throughput of IACTs and monitor their optical point spread function.
  We investigate whether muons ring images can be used as the primary optical throughput calibration method for the telescopes of the future Cherenkov Telescope Array (CTA) and find several additional systematic effects in comparison to previous works.
  To ensure that the method achieves the accuracy required by CTA, these systematic effects need to be taken into account and minor modifications to the hardware and analysis are necessary.
  We derive analytic estimates for the expected muon data rates to be used for optical throughput calibration, monitoring of the optical point spread function, with achievable statistical and systematic uncertainties, and explore the potential of muon ring images as a secondary method of camera pixel flat-fielding. }

\FullConference{36th International Cosmic Ray Conference -ICRC2019-\\
		July 24th - August 1st, 2019\\
		Madison, WI, U.S.A.}

\begin{document}

\section{Introduction}

Optical throughput calibration of Imaging Atmospheric Cherenkov Telescopes (IACTs) has been successfully carried out in the past using ring images produced by local muons~\cite{fleury,jiang:1993,vacanti,rose,rovero,puhlhofer2003,leroy,shayduk,meyermuons,goebel,humensky,hanna2008,chalmecalvet2014}. 
Given the clear advantages of the method, such as low cost, good understanding and vast experience, important collateral benefits, 
CTA has established, as a baseline, optical throughput calibration and monitoring
of its Large-, Medium- and Small-Sized-Telescopes (LST, MST  and SST), through the thorough analysis of muon rings. 

The method has been revisited in a recent paper~\cite{muonpaper} and its limitations and residual systematic uncertainties assessed. 
The task is severely complicated by the vast amount of innovative technologies foreseen for the CTA telescopes, like
dual-mirror designs~\cite{Vassiliev2007}, telescopes of very different sizes~\cite{Actis},
large fields-of-view and hence enormous camera sizes~\cite{tavernier}, or the use of silicon photomultipliers in certain camera projects~\cite{billotta,alispach}. 
The original method developed by Vacanti~et~al.~\cite{vacanti} needs  hence a list of important updates in order to achieve desired accuracies.

\section{Requirements for muon calibration}

For the muon throughput calibration to achieve the predicted 4\% accuracy~\cite{muonpaper}, a series of conditions must be met, which need to be taken into account:

\begin{description}
\item[Muon trigger:\xspace] The telescope cameras must be able to trigger on, and flag prior to transmission of data, fully contained muon rings (see e.g.~\cite{maccarone,mineo,pillera}).
\item[Hard UV blindness:\xspace] The optical elements of the Cherenkov Telescope must jointly ensure by design that wavelengths below 290~nm
  in the Cherenkov light spectrum from local muons contribute less than 5\% to the observed muon image amplitude. This requirement can actually be monitored with a
  dedicated illumination device~\cite{segreto}.
\item[Chromaticity of degradation:\xspace]  The relative degradation of the system for different wavelengths, i.e. the chromaticity of the degradation, must be assessed externally.
  Typically, such effects can  be monitored
  on yearly  time-scales~\cite{segreto},  albeit rather relaxed accuracy requirements of 10--15\% are sufficient here.
\item[Unbiased pulse integration and noise removal:\xspace] The biases introduced by pulse integration and noise removal to the muon analysis must be smaller than 1\% of the estimated ring image size. 
\item[Correction for non-active pixels:\xspace] Non-active, broken or unreliable pixels must be correctly taken into accout,
  and their impact must account for less than 0.5\% of the muon image size (see \cite{mitchellphd}).
\item[Non-spherical reflectors:\xspace]  IACT reflectors are typically not perfectly spherical, but show either an approximately hexagonal structure, or even more complicated polygons.
  Such structure of the reflector must be taken into account in the muon analysis, e.g. by line integration along the mirror surface as a function of azimuth angle~\cite{mitchell}.
\item[Plate-scale calibrations:\xspace] The muon analysis must correct for biases due to optical aberrations in the correspondence of camera coordinates to incidence angles.
\item[Finite camera focus:\xspace] The muon analysis must correct for finite camera focuses, see Sect.~\ref{sec:focus}. 
\item[Geomagnetic field effects:\xspace] The muon analysis must be corrected for geomagnetic field effects, see Sect.~\ref{sec:magnetic}.
\item[Shadow awareness:\xspace] The muon analysis must include the effect of shadows from the camera and the typical central holes in the reflector, see Sect.~\ref{sec:shadow}. 
\item[Muon simulations:\xspace] The method must be calibrated, when starting to operate each new telescope, against dedicated muon simulations, which must correctly include all
  shadowing parts of the telescope, at their respective distances to the reflector. That initial ``calibration of the method'' is necessary to account for
  residual shadows (e.g. ropes) not included in the reconstruction.
\item[Extinction of muon Cherenkov light by air molecules and aerosols:\xspace] The analysis must account for extinction of the muon Cherenkov light by air molecules
  and aerosols, see Sect.~\ref{sec:extinction}.
\item[Incidence-angle dependencies:\xspace] Dependencies of the sensitivity of the camera to light from different incidence angles must be included in the muon analysis,
  see Sect.~\ref{sec:shadow}.
\item[Mis-focused mirrors:\xspace] The muon analysis must be corrected for mis-aligned primary mirrors.
\end{description}
In the following, a small selection of important, but hitherto overlooked, details is presented.

\subsection{Finite camera focuses \label{sec:focus}}

Large IACTs focus their cameras at the mean altitude at which air shower maxima are observed, normally chosen as 10~km distance above the telescope.  

For an ideal telescope of focal length $F$, focused at infinity, i.e. with the camera placed at $z_f = F$, the Cherenkov angle of the emitted rays equals the radius of the imaged ring. For a telescope focused at a distance $x_f$, i.e. with the camera placed at $z_f$ with $1/z_f = 1/F - 1/x_f$, the rays from an on-axis muon radiating at distance $x$ will form a ring of radius:
\begin{equation}
\tan\theta_\textit{zf} = \tan\theta_c \cdot \left(1-x/x_f\right) \quad,
\label{eq:tanvf}
\end{equation}
where $\theta_c$ denotes the Cherenkov angle. 
%The values of $x=0$ and $x_f = \infty$ give $\theta_\textit{xf} = \theta_c$ as expected.
%For $x=x_f$ we get $\theta_\textit{xf} =0$; this is of course the definition of what it means for the telescope to be focused at $x_f$.
At its highest point, the Cherenkov light from a local muon is imaged into a smaller ring with radius smaller than $\theta_c$, by the time the muon impacts the mirror $(x = 0)$ the angle is the nominal value. This results in a systematic error if the ring radius is directly interpreted as the mean Cherenkov angle. 
The bias introduced on the reconstructed Cherenkov angle varies from about %: 
%\begin{equation}
%B \approx  - \frac{R}{2\theta_c} \cdot \frac{1}{x_f} \cdot E_0( \rho_R) ~.\label{eq:biasfinitefocus}
%\end{equation}
$\sim -2\%$ to $\sim -4$\%, depending on the muon energy for the MST, whereas for the LST biases can be as high as $-8\%$.

\subsection{Geomagnetic field effects \label{sec:magnetic}}

Bending of the muon trajectory in the geomagnetic field has not been considered important so far for local muons. However,
it may lead to non-negligible and asymmetric broadening of muon rings, especially for large-sized IACTs located on sites with stronger terrestrial field.
%It is not even mentioned as a source of ring broadening by Vacanti~\cite{vacanti}. 
%Using the common expression for the gyro-radius, 
A muon of momentum $p$ propagating a distance $D$ perpendicular to the field of strength $B$ will be deflected by an angle $\Delta\theta$:
\begin{equation}
\Delta\theta\quad   = ~ \left(\frac{\mathrm{d}\theta}{\mathrm{d} x}\right) \cdot x \quad %\nonumber\\
             = ~6.9\times 10^{-5} ~\mathrm{(deg/m)}
\cdot \left(\frac{10~\mathrm{GeV}}{p}\right) 
\cdot \left(\frac{B_\perp}{40~\mu\mathrm{T}}\right) 
\cdot \left(\frac{x}{\mathrm{m}}\right) \label{eq:muonbending} 
\end{equation}
\noindent
where $B_\perp$ the magnetic field component perpendicular to the muon's velocity.  Its maximum value can be assumed as $B \approx 40~\mu$T, achieved at La Palma at a zenith angle of $\sim 40^\circ$.
%The angle under which the muon Cherenkov light is imaged in the camera gets then slightly shifted by: 
%\begin{align}
%\Delta \theta_c(x,\phi)  
%                &\simeq     \cos(\varphi-\phi) \cdot \left( \frac{\mathrm{d}\theta}{\mathrm{d} x}\right) \cdot x \quad, 
%\end{align}
%where $\phi$ is the local azimuth angle of the emitted muon Cherenkov light (see \cite{muonpaper} for more details).

The bias for the reconstructed Cherenkov angle %$B(\theta_c) = (E[\Delta\theta_c])/\theta_c$
can reach up to 2\% in the worst case of a full muon ring imaged into an LST camera, under the maximum impact distance and the azimuthal component of the $B_\perp$  parallel to the azimuth component of the impact parameter.   

%Geomagnetic field effects can be studied straight-forwardly  by comparing data from pointings towards the South (where the perpendicular magnetic field is strongest) and the North (where it is weakest, almost reaching zero, (see \cite{commichau2008}) in the case of La Palma. 

\subsection{Non-negligible inclination angles and camera shadows \label{sec:shadow}}

Muon inclination angles $\upsilon$ were so far limited to $\upsilon \lesssim 1^\circ$ for field-of-view cameras $\lesssim 5^\circ$,
with associated effects on the throughput calibration of only $ \lesssim 0.05\%$.
However, wide-field cameras for CTA with up to 10$^\circ$ field-of-view can  regularly obtain muon images with ring centers up to $3.5^\circ$ from the camera center,
resulting in an effect on the modulated signal along the ring of $\lesssim 0.3$\%. 

The situation becomes more complicated when central shadowing objects, such as the camera, are considered.
%Traditionally, relatively small cameras (with small focal ratios and hence small  dimensions compared to the reflector) have been employed by IACTs.
%In these cases, the shadowing function $\Theta(\vec{\rho},\vec{\psi_c})$ can be approximated by Eq.~\ref{eq:Thetaapprox}.
For telescopes with a  focal ratio of $f$, the shadow produced by the camera moves with inclination angle approximately as $ 
%\begin{align}
\Delta \approx f \cdot \upsilon  
%              & \approx 0.07 \cdot \rho_R \quad \textrm{for:}\quad\psic = 2^\circ 
%\quad, 
%\end{align}
%\noindent
$, 
slightly exceeding the intrinsic resolution, with which the muon impact distance can be obtained.
%The apparent shift of the camera shadow can, however, be included in the reconstruction of the impact parameters and optical bandwidth using Eq.~\ref{eq:rho_moved}. 

\begin{figure}
\centering
%%% figure produced by 'pyscripts/ntot.py'
\includegraphics[width=0.63\linewidth]{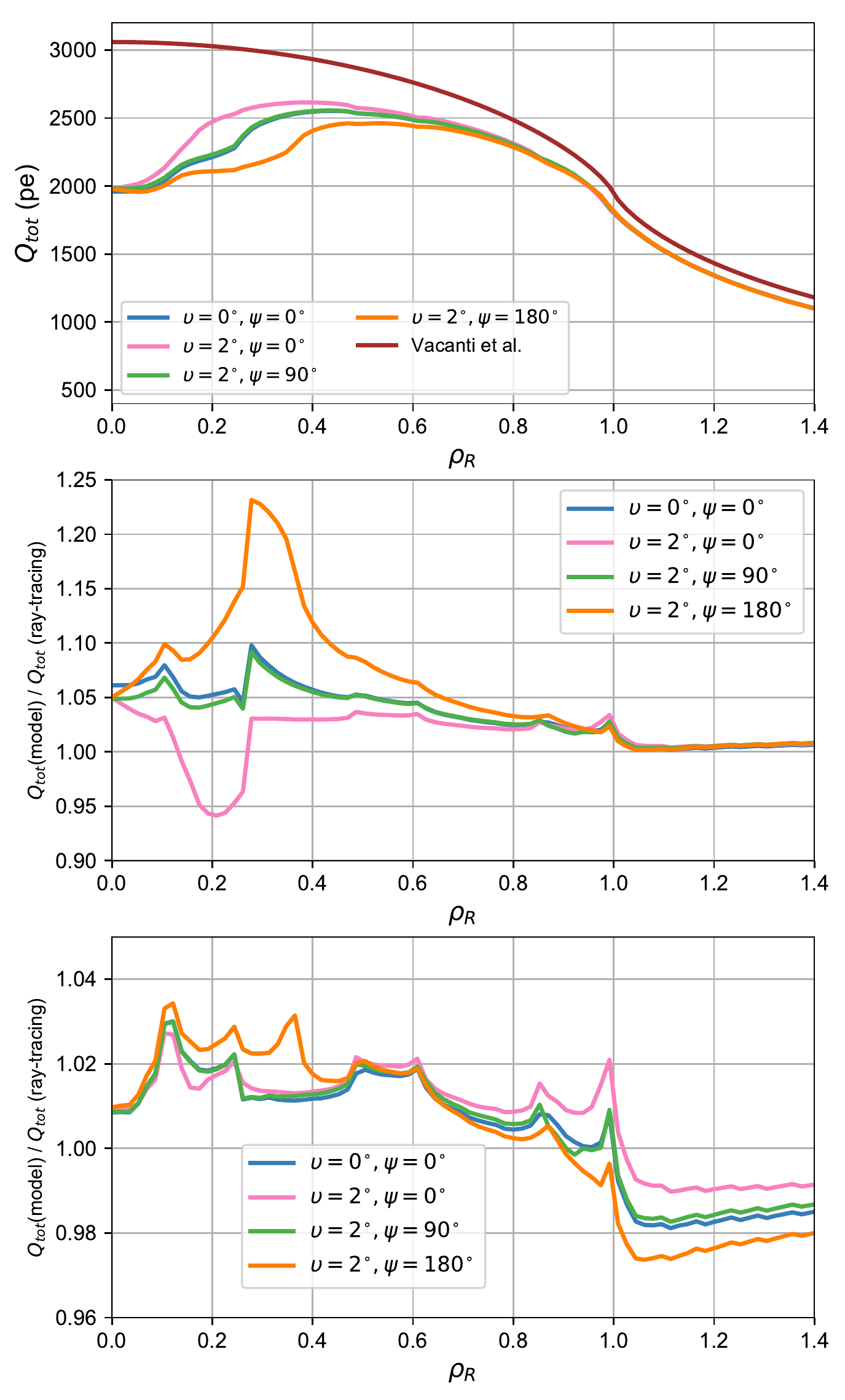}
\caption{Expected effect of shadowing of the MST's square camera on the total amount of photons collected (top).
  In full brown, Vacanti's solution~\cite{vacanti} is plotted without considering the shadow.  %, in dashed brown a roundish analytic camera model has been subtracted from the previous one.
  %The dashed-dotted line uses instead an analytical solution for the shadow of a square-shaped camera and uses Eqs.~\ref{eq:Dshadow} and~\ref{eq:Dhole}.
  The full blue lines show the true value for an on-axis muon, and the orange, green and pink lines show instead the case for a 2$^\circ$ inclined muon,
  with azimuthal incidence angles of $0^\circ, 90^\circ$, and $180^\circ$.
  %A $\phi$ value of 180$^\circ$ has been used for all data points shown.
  The center plot shows the relative values of the latter cases, compared to the analytical roundish camera model
  and the bottom plot with the analytical square camera model and the improved shadow models.
  Note the different y-axis scales of the center and bottom plots.
  With the full shadow model,
  a peak-to-peak difference in modelled and simulated photon yield between the investigated impact distances and muon inclination angles of only about 4\% is observed,
  if fully contained rings are considered. 
\label{fig:totphotonsMST}}
\end{figure}

Furthermore, the square shape of some camera designs (like the MST and the LST) introduces dependencies of the reconstructed optical throughput on the muon inclination angle,
see Fig.~\ref{fig:totphotonsMST}.
The effect of such a quadratic shape can nevertheless be included in the analytical model, as well as the additional shadow of the central hole.

\subsection{Extinction of muon Cherenkov light \label{sec:extinction}}

A photon of energy $\epsilon$ emitted at a distance $x$ from the telescope suffers molecular and aerosol extinction. 
One can describe the atmospheric transmission for the photon as: 
\begin{equation}
T(\epsilon,x;\vartheta)  = \exp \big(  - \int_0^{x\cdot \cos\vartheta} \!\! \left[ \alpha_\mathrm{mol}(\epsilon,h) + \alpha_\mathrm{aer}(\epsilon,h) \right] ~\mathrm{d} h / \cos\vartheta\big)  \quad,
\end{equation}
\noindent
where $\alpha_\mathrm{mol}$ and $\alpha_\mathrm{aer}$ are the volume extinction coefficients from molecular and aerosol extinction at an altitude $h$, respectively,
and the telescope itself points to the sky under a zenith angle $\vartheta$.

Large telescopes are more prone to variations of the atmospheric extinction because they observe local muons emitting from higher altitudes. 
We simulated typical variations of the molecular profile at La Palma~\cite{munar} and found peak-to-peak variations of the molecular transmission smaller than 0.2\%.
Nevertheless, aerosol transmission can be affected by dust intrusions, particularly at La Palma.  In the case of strong aerosol densities,
very close to the ground, a correction of $\lesssim 3$\% 
needs to be applied. This problem can be either circumvented by adequate data selection, 
or by applying a bias correction,
which can be calculated once the aerosol profile is assessed. %Such a correction is expected to take place only rarely, when the observatory operates under non-optimal conditions.

\section{Additional use of muon calibration}

Contrary to the Cherenkov light from gamma-ray induced showers, Cherenkov light from local muons does not illuminate the reflector uniformly, 
but is instead concentrated at the muon impact point.  Moreover, the expected fluence of Cherenkov light at  each camera pixel can be predicted  to  good accuracy~\cite{muonpaper}. 
For these  reasons,  muon calibration delivers additional calibration information, which can be exploited for alternative calibrations and used for monitoring:

\begin{description}
\item[Pixel-wise charge calibration:\xspace] Flat-fielding coefficients from muon analysis can be used to monitor and improve the accuracy
of the coefficients obtained through the typical camera  flat-fielding.
\item[Pixel-wise time calibration:\xspace] Time offset coefficients from muon analysis can be used to monitor and improve the accuracy of the coefficients
  obtained with the flat-fielding procedure.
\item[Optical PSF monitoring:\xspace] Estimates of the optical point spread function (PSF) of each telescope from muon analysis can be used to  monitor
  and improve its accuracy.
\item[Degradation of individual mirrors:\xspace] Estimates of the relative reflectance of each mirror with respect to the telescope average
can be obtained from muon analysis. 
\end{description}

Table~\ref{tab:ratelistfinal} lists the expected achievable  rates, with which each monitoring point can be obtained to a precision comparable with the expected accuracy of the same monitoring point (see~\cite{muonpaper} for details).
One can see that the needed monitoring times are typically shorter than the duration of  one science observation run.

\begin{table}[h!]
\centering
  \renewcommand{\arraystretch}{0.5}
\begin{tabular}{lccccccc} \toprule \addlinespace[0.2cm]
 & \multicolumn{2}{c}{LST}  &  \multicolumn{2}{c}{MST} &  \multicolumn{2}{c}{SST}\\ \addlinespace[0.2cm]
 & $\vartheta=0^\circ$ & $\vartheta=60^\circ$  & $\vartheta=0^\circ$ & $\vartheta=60^\circ$ & $\vartheta=0^\circ$ & $\vartheta=60^\circ$  \\ \addlinespace[0.1cm] \midrule
%\rowcolor{tablightblue} Maximum impact distance $\rho_\mathrm{max}$ (m) & \multicolumn{2}{c|}{11.8}  & \multicolumn{2}{c}{4.9}   \\
%\rowcolor{tablightblue} Maximum muon angle $\upsilon_\mathrm{max}$ ($^\circ$) & \multicolumn{2}{c|}{0.8} &  \multicolumn{2}{c}{2.5}    \\
% \midrule
& \multicolumn{6}{c}{\bf Determination of optical throughput } \\\addlinespace[0.2cm]
 Rate (Hz)  &  1.8  & 0.5  & 5 & 1.4   & 0.8  & 0.2  \\ \addlinespace[0.2cm]
 Monitoring time & 14~s  & 48~s &  5~s   & 18~s  & 28~min & 18~min   \\\addlinespace[0.3cm] \midrule \addlinespace[0.1cm]
% \midrule
& \multicolumn{6}{c}{\bf Determination of optical throughput (per pixel)  } \\ \addlinespace[0.2cm]
 Rate (Hz)  & 0.07   & 0.02   &  0.10   & 0.03 & 0.02 & 0.006    \\ \addlinespace[0.2cm]
 Monitoring time & 6~min & 21~min  &  4~min & 13~min & 2.3~h & 7.6~h  \\\addlinespace[0.3cm]  \midrule  \addlinespace[0.1cm]
%\midrule
& \multicolumn{6}{c}{\bf Determination of time offset (per pixel)  } \\\addlinespace[0.2cm]
 Rate (Hz)  & 0.1  & 0.03 & 0.2 & 0.04 & 0.03 & 0.008 \\ \addlinespace[0.2cm]
Monitoring time &  10~s & 36~s & 6~s & 23~s & 36~s & 2.2~min  \\\addlinespace[0.3cm] \midrule \addlinespace[0.1cm]
%\midrule
& \multicolumn{6}{c}{\bf Determination of optical PSF  } \\\addlinespace[0.2cm]
 Rate (Hz) & 5 & 1.6     & 8    & 2.5 & 2.4 & 0.7   \\\addlinespace[0.2cm]
 Monitoring time &  9~s & 25~s & 5~s  & 16~s & 17~s & 1~min  \\\addlinespace[0.1cm]  
\bottomrule
\end{tabular}  
\caption{\label{tab:ratelistfinal}Expected rates of high quality, fully contained muon ring images for calibration for the proposed telescopes of CTA, and the corresponding monitoring time scales. %\tem{It is not clear to me where the maximum angles for LST and MST comes from (???Fov/2-1.3 ???).
%\tem{The table presents some discrepancies wrt the text, i.e: 50 usable events for LST and MST are expected for 7percent RMS; in the next there is no info about RMS for SC-MST; how column ERgood at 60deg is computed? this is not only a reduction due to cos2Theta; in the second column, ASTRI time should be 33 min, not 40 min.}
}
\end{table}

\section{Conclusions}

Using algorithms from Vacanti et~al.~\cite{vacanti}, 
several authors have claimed a theoretical accuracy of as good as $\sim$2\% for the optical throughput calibration of IACTs using images from local muons.
To ensure that the method achieves the accuracy required by CTA, we revisited all documented algorithms and systematic effects~\cite{muonpaper}
and found several effects, which have not been taken into account so far, or which appear non-negligible for large mirrors, or large square-shaped cameras,
or the dual-mirror telescope designs proposed for CTA. 

Most of these effects can be corrected by a careful analysis. Nevertheless, a few systematic effects
%, particularly those related with the differences in the observed spectrum of Cherenkov light received from local muons and distance gamma-ray showers, 
will require dedicated studies or even adaptation in the design of the hardware employed for the CTA telescopes to achieve the desired accuracy. %Other effects require careful understanding or even modifications in the design of the hardware employed for the CTA telescopes. 
These effects are mostly related to the differences in the observed spectrum of Cherenkov light between local muons and distant gamma-ray showers.
%The conversion of measured optical bandwidth with the muon method to its equivalent for the telescopes' response to Cherenkov light from gamma-ray showers, to account for wavelength-dependent degradation, requires hardware that efficiently cuts the camera response to photon 
%energies beyond about 4.3~eV. 
%wavelengths below about 290~nm.
%Finally, close comparison with Monte Carlo simulated telescope response to Cherenkov light emitted by local muons is necessary at the beginning of operation, to correct for residual shadows and spaces between individual mirrors. 
%Deviations from the perfect spherical geometry assumed by the analytical formulae can be .  

We find that after incorporating all effects, the optical throughput of the full telescope can be determined with an accuracy of better than 4\% (LST and MST) and 5\% (SST).
%This is considerably better than other proposed methods so far~\citep{gaug2014,Brown:2018,Stefanik:2018} and only slightly worse than the cross-calibration scheme proposed by~\citet{mitchell2015}, which allows, however, only for a relative calibration between telescopes or telescope types. Other direct methods~\citep{segreto2016} perform even better  and show more flexibility, but require a dedicated setup and cannot be carried out \textit{at the same time as the telescopes perform science observations}. The muon calibration method is therefore the first option for  \textit{online} and \textit{offline} monitoring of the optical bandwidth of all CTA telescopes~\citep{gaugSPIE2014}.

Muon calibration allows one to monitor additionally, and without much effort the  flat-fielding of the camera,
and to provide the relative time offsets of each pixel.
%, at least for the large 
%telescopes, but could also be tried for the SSTs. 
%An automatic monitoring algorithm shall be implemented in the off-site analysis. 
Similarly, differences in reflectance among the mirrors can be detected by plotting the retrieved optical bandwidth as a function of reconstructed impact point.% (Eq.~\ref{eq:vecrho}). 
%This procedure should work at least for the SSTs and the MSTs. Because of the larger mirror area covered by LST muon light, further investigation is necessary to find out whether such an 
%approach is also useful for these telescopes.

Finally, contrary to the starlight,
the light from local muons gets registered within sub-nanosecond time windows,
which allows one to greatly reduce residual backgrounds from the night sky and unresolved stars and monitor the optical point spread function of each telescope.
%The stronger concentration of Cherenkov light from local muons close to their impact points may even enable  point-spread monitoring resolved to individual mirror facets or groups of mirrors. 

%We expect that the combination of these methods will provide a robust estimate of the average combined telescope throughput and camera gain to an accuracy of better than 5\%. 

\acknowledgments
This work was conducted in the context of the CTA Central Calibration Facilities Working Group. 
We gratefully acknowledge financial support from the agencies and organizations listed here: \url{http://www.cta-observatory.org/consortium_acknowledgments}
This proceeding has gone through internal review by the CTA Consortium.


\begin{thebibliography}{99}
\bibitem{fleury}
P. {Fleury} et al., \emph{{\v C}erenkov ring
  images of cosmic ray muons}. Proc. 22$^\mathrm{nd}$
  ICRC, Dublin {\bf 2} 595, 1991.
\bibitem{jiang:1993}
{Jiang}, Y., {Fleury}, P., {Lewis}, A.~D., et~al. \emph{Absolute Calibration
  of an Atmospheric Cherenkov Telescope Using Muon Ring Images}. Proc. 23$^\mathrm{rd}$
  ICRC {\bf 4} 662, 1993.
\bibitem{vacanti}
{Vacanti}, G., {Fleury}, P., {Jiang}, Y., et~al. \emph{Muon ring images with
  an atmospheric {\v C}erenkov telescope} \emph{Astrop. Phys.} {\bf 2}, 1--11, 1994. 
\bibitem{rose}
{Rose}, H.~J. \emph{Cherenkov Telescope Calibration using Muon Ring Images}.
  Proc. 24$^\mathrm{th}$ ICRC {\bf 3} 464, 1995. 

\bibitem{rovero}
{Rovero}, A.~C., {Buckley}, J.~H., {Fleury}, P., et~al. \emph{Calibration of
  the Whipple atmospheric {\v C}herenkov telescope} \emph{Astrop. Phys.} {\bf 5},
  27--34, 1996.

\bibitem{puhlhofer2003}
{P{\"u}hlhofer}, G., {Bolz}, O., {G{\"o}tting}, N., et~al. \emph{The
  technical performance of the HEGRA system of imaging air Cherenkov
  telescopes} \emph{Astrop. Phys.} {\bf 20}, 267--291, 2003. 

\bibitem{leroy}
{Leroy}, N., {Bolz}, O., {Guy}, J., et~al. \emph{Calibration Results for the
  First Two H.E.S.S. Array Telescopes} Proc. 28$^\mathrm{th}$
  ICRC {\bf 5} 2895, 2003. 


\bibitem{shayduk}
{Shayduk}, M., {Kalekin}, O., {Mase}, K., {Pavel}, N., {MAGIC Collaboration}
  \emph{Calibration of the MAGIC Telescope Using Muon Ring Images}, 
  Proc. 28$^\mathrm{th}$ ICRC, Tsukuba {\bf 5} 2951, 2003. 

\bibitem{meyermuons}
{Meyer}, M., et~al. \emph{Analysis of muon events recorded with the MAGIC
  telescope}. High Energy Gamma-Ray Astronomy  {\bf 745} 774--778, 2005. 


  \bibitem{goebel}
{Goebel}, F., {Mase}, K., {Meyer}, M., et~al. \emph{Absolute energy scale
  calibration of the MAGIC telescope using muon images}. Proc.
  29$^\mathrm{th}$ ICRC {\bf 5} 179, 2005. 

  \bibitem{humensky}
{Humensky}, T.~B.  
  \emph{Calibration of VERITAS Telescope 1 via Muons}. Proc. 29$^\mathrm{th}$
  ICRC, Pune,  2005 [{\tt ArXiv:astro-ph/0507449}]

  \bibitem{hanna2008}
{Hanna}, D. \emph{Calibration Techniques for VERITAS}. Proc.
  30$^\mathrm{th}$ ICRC, Merida {\bf 3} 
  1417--1420,  2008. 


\bibitem{chalmecalvet2014}
 {Chalme-Calvet}, R., {de Naurois}, M., {Tavernet}, J.-P. {for the
  H.~E.~S.~S.~Collaboration} \emph{Muon efficiency of the H.E.S.S.
  telescope}. Proc. AtmoHEAD Conf., Saclay 2014.
  [{\tt arXiv:1403.4550}]

\bibitem{muonpaper}
  M.~Gaug, S.~Fegan, A.~M.~W.~Mitchell, M.~C.~Maccarone, T.~Mineo, A.~Okumura,
  \emph{Using Muon Rings for the Calibration of the Cherenkov Telescope Array: A Systematic Review of the Method and Its Potential Accuracy},
  \emph{Astrophys. J. Suppl.} {\bf 243} 11, 2019 [{\tt arXiv:1907.04375}].

\bibitem{Vassiliev2007}
  V.~{Vassiliev}, S.~{Fegan} and P.~{Brousseau}, \emph{Wide field aplanatic two-mirror telescopes for ground-based {$\gamma$}-ray astronomy},
	\emph{Astroparticle Physics}  {\bf 28} 10-27, 2007.

\bibitem{Actis} 
	{The CTA Consortium:~Actis, M. et~al.}, 
	\emph{Design concepts for the Cherenkov Telescope Array CTA: an advanced facility for ground-based high-energy gamma-ray astronomy}, 
	\emph{Experimental Astronomy} {\bf 32} 193-316, 2011 [{\tt arXiv:1008.3703}]

\bibitem{tavernier}
  T.~Tavernier et~al., \emph{Status and performance results from NectarCam, a camera for CTA MST's},  these proceedings.

\bibitem{billotta}
	{Billotta}, S. et~al. \emph{SiPM Detectors for the ASTRI project in the framework of the Cherenkov Telescope Array}, 
        \emph{Proc. SPIE Astronomical Telescopes + Instrumentation} {\bf 9154} 91541R, 2014. 

\bibitem{alispach}
  C.~Alispach et~al., \emph{Calibration strategy for the next generation of SiPM cameras},  these proceedings.

\bibitem{maccarone}
  M.~C.~Maccarone et al., \emph{Pre-selecting muon events in the camera server of the ASTRI telescopes for the Cherenkov Telescope Array}, \emph{Proc. SPIE} {\bf 9913} 991370, 2016. 
  
\bibitem{mineo}
  T.~Mineo et~al., \emph{Muon calibration of the ASTRI-Horn telescope: preliminary results}, these proceedings.

\bibitem{pillera}
  R.~Pillera et~al., \emph{Muon tagging on the BEE of CHEC-S -- a compact high-energy camera for CTA}, these proceedings.

\bibitem{segreto}
  A.~Segreto et~al., \emph{The absolute calibration strategy of the ASTRI SST-2M telescope proposed for the Cherenkov Telescope Array and its external ground-based illumination system},
        \emph{Proc. SPIE Ground-based and Airborne Telescopes VI} {\bf 9906} 99063S, 2016. 

\bibitem{mitchellphd}
  A.~{Mitchell}, \emph{Optical Efficiency Calibration for Inhomogeneous IACT Arrays and a Detailed Study of the Highly Extended Pulsar Wind Nebula HESS J1825-137}, 
  PhD thesis, University Heidelberg, 2016.
        
\bibitem{mitchell}
  A. {Mitchell}, V.~{Marandon}, and D. {Parsons} \emph{A Generic Algorithm for IACT Optical Efficiency Calibration using Muons} 
   \emph{Proc. 34$^\mathrm{th}$ ICRC,  The Hague} {\bf 236} 1, 2016 [{\tt arXiv:1509.042587}]

 \bibitem{munar}
   P. {Munar-Adrover} and M. {Gaug}, \emph{Studying molecular profiles above the Cherenkov Telescope Array sites}, 
   \emph{EPJ} {\bf 197} 01002, 2019. 
   
   
\end{thebibliography}
\end{document}